\documentclass[ALICE,manyauthors]{cernphprep}
\usepackage[comma,square,numbers,sort&compress]{natbib}
\usepackage{hyperref}
\usepackage{lineno}
\usepackage{xspace}
\usepackage{upgreek}
\usepackage{xcolor}
\usepackage{multirow}
\usepackage{amsmath}
\usepackage{euscript}
\usepackage{colortbl}
\usepackage[T1]{fontenc}
\begin{document}
%%%%%%%%%%%%%%%%%%%%%%%%%%%%%%%%%%%%%%%%%%%%%%%%%%
% These are some new commands that may be useful 
% for paper writing in general. If other newcommands
% are needed for your specific paper, please feel 
% free to add here. 
%
% The currently available commands are organized in: 
% 1) Systems
% 2) Quantities
% 3) Energies and units
% 4) Detectors
% 5) particle species 
%%%%%%%%%%%%%%%%%%%%%%%%%%%%%%%%%%%%%%%%%%%%%%%%%%

% 1) SYSTEMS 
\newcommand{\pp}           {pp\xspace}
\newcommand{\ppbar}        {\mbox{$\mathrm {p\overline{p}}$}\xspace}
\newcommand{\XeXe}         {\mbox{Xe--Xe}\xspace}
\newcommand{\PbPb}         {\mbox{Pb--Pb}\xspace}
\newcommand{\pA}           {\mbox{pA}\xspace}
\newcommand{\pPb}          {\mbox{p--Pb}\xspace}
\newcommand{\AuAu}         {\mbox{Au--Au}\xspace}
\newcommand{\dAu}          {\mbox{d--Au}\xspace}

% 2) QUANTITIES 
\newcommand{\s}            {\ensuremath{\sqrt{s}}\xspace}
\newcommand{\snn}          {\ensuremath{\sqrt{s_{\mathrm{NN}}}}\xspace}
\newcommand{\pt}           {\ensuremath{p_{\rm T}}\xspace}
\newcommand{\ks}           {\ensuremath{k^*}\xspace}
\newcommand{\meanpt}       {$\langle p_{\mathrm{T}}\rangle$\xspace}
\newcommand{\ycms}         {\ensuremath{y_{\rm CMS}}\xspace}
\newcommand{\ylab}         {\ensuremath{y_{\rm lab}}\xspace}
\newcommand{\etarange}[1]  {\mbox{$\left | \eta \right |~<~#1$}}
\newcommand{\yrange}[1]    {\mbox{$\left | y \right |~<~#1$}}
\newcommand{\dndy}         {\ensuremath{\mathrm{d}N_\mathrm{ch}/\mathrm{d}y}\xspace}
\newcommand{\dndeta}       {\ensuremath{\mathrm{d}N_\mathrm{ch}/\mathrm{d}\eta}\xspace}
\newcommand{\avdndeta}     {\ensuremath{\langle\dndeta\rangle}\xspace}
\newcommand{\dNdy}         {\ensuremath{\mathrm{d}N_\mathrm{ch}/\mathrm{d}y}\xspace}
\newcommand{\Npart}        {\ensuremath{N_\mathrm{part}}\xspace}
\newcommand{\Ncoll}        {\ensuremath{N_\mathrm{coll}}\xspace}
\newcommand{\dEdx}         {\ensuremath{\textrm{d}E/\textrm{d}x}\xspace}
\newcommand{\RpPb}         {\ensuremath{R_{\rm pPb}}\xspace}

% 3) ENERGIES, UNITS
\newcommand{\nineH}        {$\sqrt{s}~=~0.9$~Te\kern-.1emV\xspace}
\newcommand{\seven}        {$\sqrt{s}~=~7$~Te\kern-.1emV\xspace}
\newcommand{\twoH}         {$\sqrt{s}~=~0.2$~Te\kern-.1emV\xspace}
\newcommand{\twosevensix}  {$\sqrt{s}~=~2.76$~Te\kern-.1emV\xspace}
\newcommand{\five}         {$\sqrt{s}~=~5.02$~Te\kern-.1emV\xspace}
\newcommand{\twosevensixnn}{$\sqrt{s_{\mathrm{NN}}}~=~2.76$~Te\kern-.1emV\xspace}
\newcommand{\fivenn}       {$\sqrt{s_{\mathrm{NN}}}~=~5.02$~Te\kern-.1emV\xspace}
\newcommand{\LT}           {L{\'e}vy-Tsallis\xspace}
\newcommand{\GeVc}         {Ge\kern-.1emV/$c$\xspace}
\newcommand{\MeVc}         {Me\kern-.1emV/$c$\xspace}
\newcommand{\TeV}          {Te\kern-.1emV\xspace}
\newcommand{\GeV}          {Ge\kern-.1emV\xspace}
\newcommand{\MeV}          {Me\kern-.1emV\xspace}
\newcommand{\GeVmass}      {Ge\kern-.2emV/$c^2$\xspace}
\newcommand{\MeVmass}      {Me\kern-.2emV/$c^2$\xspace}
\newcommand{\lumi}         {\ensuremath{\mathcal{L}}\xspace}

\newcommand{\mevc}         {{\rm{MeV}}/c\xspace}

% 4) DETECTORS 
\newcommand{\ITS}          {\rm{ITS}\xspace}
\newcommand{\TOF}          {\rm{TOF}\xspace}
\newcommand{\ZDC}          {\rm{ZDC}\xspace}
\newcommand{\ZDCs}         {\rm{ZDCs}\xspace}
\newcommand{\ZNA}          {\rm{ZNA}\xspace}
\newcommand{\ZNC}          {\rm{ZNC}\xspace}
\newcommand{\SPD}          {\rm{SPD}\xspace}
\newcommand{\SDD}          {\rm{SDD}\xspace}
\newcommand{\SSD}          {\rm{SSD}\xspace}
\newcommand{\TPC}          {\rm{TPC}\xspace}
\newcommand{\TRD}          {\rm{TRD}\xspace}
\newcommand{\VZERO}        {\rm{V0}\xspace}
\newcommand{\VZEROA}       {\rm{V0A}\xspace}
\newcommand{\VZEROC}       {\rm{V0C}\xspace}
\newcommand{\Vdecay} 	   {\ensuremath{V^{0}}\xspace}

% 4) PARTICLE SPECIES 
\newcommand{\ee}           {\ensuremath{e^{+}e^{-}}} 
\newcommand{\pip}          {\ensuremath{\pi^{+}}\xspace}
\newcommand{\pim}          {\ensuremath{\pi^{-}}\xspace}
\newcommand{\kap}          {\ensuremath{\rm{K}^{+}}\xspace}
\newcommand{\kam}          {\ensuremath{\rm{K}^{-}}\xspace}
\newcommand{\p}         {\ensuremath{\rm{p}}\xspace}
\newcommand{\n}         {\ensuremath{\rm{n}}\xspace}
\newcommand{\pbar}         {\ensuremath{\rm\overline{p}}\xspace}
\newcommand{\kzero}        {\ensuremath{{\rm K}^{0}_{\rm{S}}}\xspace}
\newcommand{\lmb}          {\ensuremath{\Lambda}\xspace}
\newcommand{\almb}         {\ensuremath{\overline{\Lambda}}\xspace}
\newcommand{\Om}           {\ensuremath{\Omega^-}\xspace}
\newcommand{\Mo}           {\ensuremath{\overline{\Omega}^+}\xspace}
\newcommand{\X}            {\ensuremath{\Xi^-}\xspace}
\newcommand{\Ix}           {\ensuremath{\overline{\Xi}^+}\xspace}
\newcommand{\Xis}          {\ensuremath{\Xi^{\pm}}\xspace}
\newcommand{\Oms}          {\ensuremath{\Omega^{\pm}}\xspace}
\newcommand{\degree}       {\ensuremath{^{\rm o}}\xspace}
%%%%%%%%%%%%%%%%%%%%%%%%%%

%%%%%%%%%%%%%%%  Title page %%%%%%%%%%%%%%%%%%%%%%%%
\begin{titlepage}
% the dates below correspond to CERN approval
% please don't touch: EB chairs will take care
\PHyear{2021}       % required, will be obtained from CERN
\PHnumber{080}      % required, will be obtained from CERN
\PHdate{10 May}  % required, will be obtained from CERN
%%%%%%%%%%%%%%%%%%%%%%%%%%%%%%%%%%%%%%%%%%%%%%%%%%%%

%%% Put your own title + short title here:
\title{Kaon--proton strong interaction at low relative momentum via femtoscopy in Pb--Pb collisions at the LHC}
\ShortTitle{Kaon--proton scattering in Pb--Pb collisions at the LHC}   % appears on left page headers

%%% Do not change the next lines
\Collaboration{ALICE Collaboration\thanks{See Appendix~\ref{app:collab} for the list of collaboration members}}
\ShortAuthor{ALICE Collaboration} % appears on right page headers, do not change

\begin{abstract}
In quantum scattering processes between two particles, aspects characterizing the strong and Coulomb forces can be observed in kinematic distributions of the particle pairs. 
The sensitivity to the interaction potential reaches a maximum at low relative momentum and vanishing distance between the two particles. Ultrarelativistic heavy-ion collisions at the LHC provide an abundant source of many hadron species and can be employed as a measurement method of scattering parameters that is complementary to scattering experiments.
This study confirms that momentum correlations of particles produced in Pb--Pb collisions at the LHC provide an accurate measurement of  kaon--proton scattering parameters at low relative momentum, allowing precise access to the $\kam p\rightarrow\kam p$ process. 
This work also validates the femtoscopic measurement in ultrarelativistic heavy-ion collisions as an alternative to scattering experiments and a complementary tool to the study of exotic atoms with comparable precision.
In this work, the first femtoscopic measurement of momentum correlations of $\kam\p\ (\kap\overline{p})$ and $\kap\p\ (\kam\overline{p})$ pairs in Pb--Pb collisions at centre-of-mass energy per nucleon pair of $\snn=5.02$~TeV registered by the ALICE experiment is reported. 
The components of the $\kam\p$ complex scattering length are extracted and found to be $\Re f_0=-0.91\pm~{0.03}$(stat)$^{+0.17}_{-0.03}$(syst) and $\Im f_0 = 0.92\pm~{0.05}$(stat)$^{+0.12}_{-0.33}$(syst). The results are compared with chiral effective field theory predictions as well as with existing data from dedicated scattering and exotic kaonic atom experiments. 
\end{abstract}
\end{titlepage}

\setcounter{page}{2} %please do not remove this line

%%%%%%%%%%%%%%%%%%%%%%%%%%%%%%%%%%%%%%%%%%%%%%%%%%%%
%%%%%%% introduction  %%%%%%%%%%%%%%%%%%%%%%%%%%%%%%
%%%%%%%%%%%%%%%%%%%%%%%%%%%%%%%%%%%%%%%%%%%%%%%%%%%%
\section{Introduction}
\label{sec:intro}

\begin{figure}[h!]
    \centering
    \includegraphics[width=0.95\textwidth]{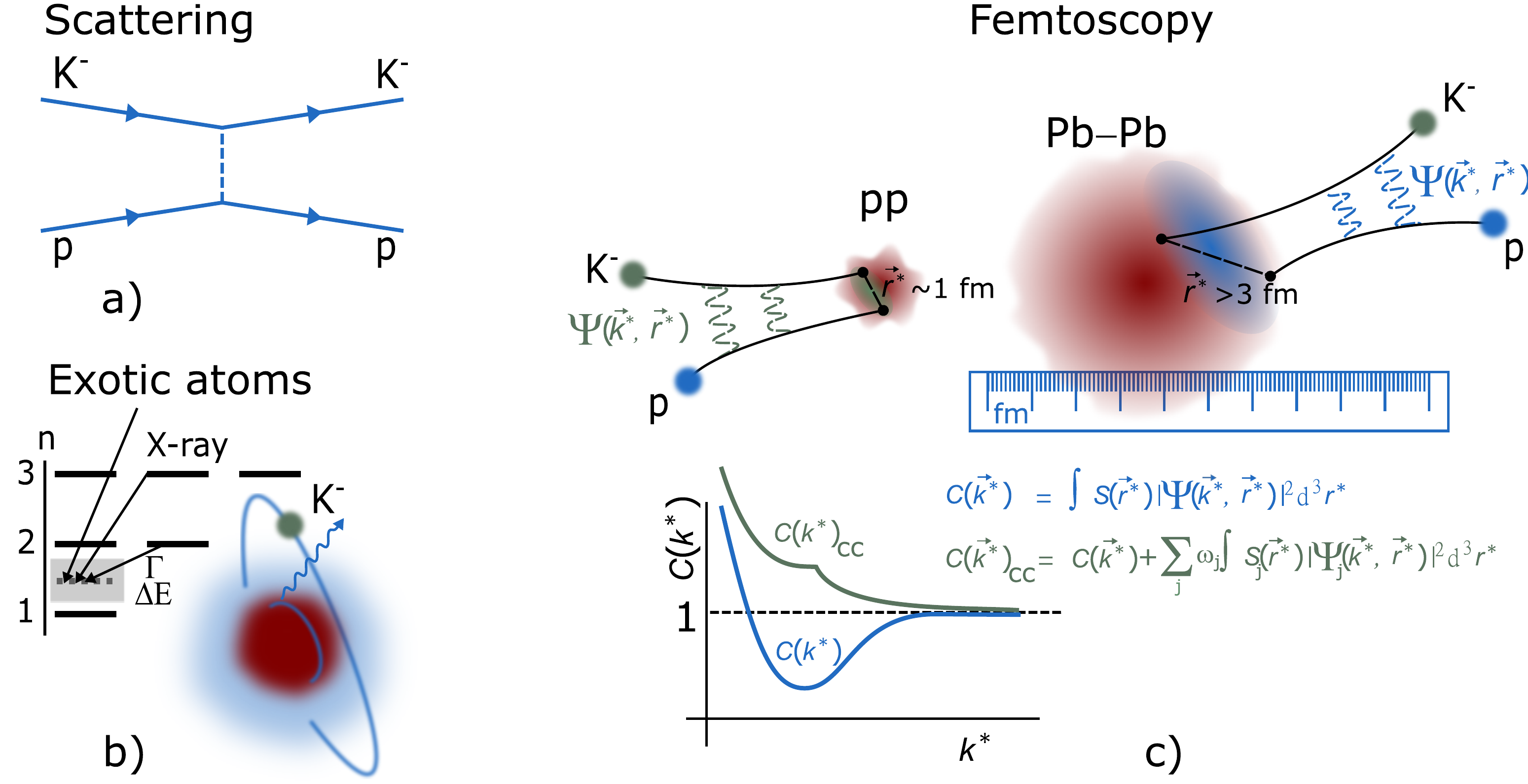}
    \caption{Schematic illustration of available experimental techniques for measuring the interaction among hadrons:  a) scattering experiments, b) measurements of energy shifts from the X-ray de-excitation spectrum of exotic atoms, c) femtoscopy in small collision systems (pp) with coupled channel effects shown in green, and in large collision systems (Pb--Pb) with vanishing coupled channel contributions shown in blue.}
    \label{fig:scheme}
\end{figure}

The study of kaon--nucleon (KN) and antikaon--nucleon ($\overline{\mathrm{K}}$N) interactions is essential to understand  low-energy quantum chromodynamics (QCD). 
The Goldstone boson nature of the kaon and its role as an active degree of freedom within effective field theories are used to describe processes within the chiral SU(3) dynamics at low energies. Since perturbation theory is not applicable in this energy regime, experimental data are essential to constrain the currently available effective theories ~\cite{Petrov:2016azi}.
In general, three experimental techniques can be used to access hadron--hadron interaction, such as $\rm \overline{K}N$, which will be discussed below. Their schematic illustration is shown in Fig.~\ref{fig:scheme}. 

A direct measurement of the $\rm \overline{K}N$ interaction was first possible in scattering experiments with secondary $\kam$ beams impinging on hydrogen targets. Such experiments allow the precise control of both the initial (incoming) and final (outgoing) states, see Fig.~\ref{fig:scheme} a. Furthermore, polarised beams and targets can be used~\cite{Leader:2001gr}. However, scattering measurements can only be performed for pairs of charged and long-lived particles, because of the technical challenges of creating beams of other particle species, especially involving strangeness or antimatter~\cite{Wiedemann:2034423}. Finally, attaining zero relative momentum for strange hadrons, such as kaons, is also challenging. Current $\kam p$ measurements can only reach a kaon momentum of approximately $100$~MeV/$c$ with respect to the proton at rest~\cite{Humphrey:1962zz,Sakitt:1965kh,Kim:1965zzd,Kim:1967zze,Kittel:1966zz,Evans:1983hz,Ciborowski:1982et}. 
Therefore, the present accuracy of the $\kam{\rm p}$ cross sections and branching ratios at threshold does not sufficiently constrain the scattering amplitudes. 

Another way of studying the low-energy $\rm \overline{K}N$ interaction is by observing the shifts and widths of energy levels in exotic kaonic atoms~\cite{Gasser:2009wf}, where a negative kaon replaces an electron, see Fig.~\ref{fig:scheme} b. In particular, kaonic hydrogen and kaonic deuterium can be formed in dedicated experimental setups and measurements of their X-ray deexcitation spectra provide a link to the $\kam p$ strong interaction at zero relative momentum~\cite{Curceanu:2019uph}. When the negative hadron, such as $\kam$, transits between the lowest energy states, its wave function overlaps with the nucleus and the presence of the strong force modifies the energy levels and their widths due to coupling to intermediate states before absorption~\cite{BATTY1997385,Hori:2719093}.  
The measurement of the energy shift and width of the 1S state of the kaonic hydrogen line can be connected to the $\kam$p scattering length ($f_0$) through the Trueman--Deser formula~\cite{Deser:1954vq,Meissner:2004jr}. However, it requires model-dependent corrections to account for the presence of the isospin-breaking couplings and the Coulomb interaction~\cite{Meissner:2004jr}. The scattering length is a complex number related to both the elastic and inelastic cross sections of the interaction. 
The current benchmark result is from chiral effective field theory ($\chi^{\rm EFT}$ calculations)~\cite{Ikeda:2011pi,Ikeda:2012au} anchored to the kaonic hydrogen data~\cite{Bazzi:2011zj,Bazzi:2012eq} collected by the SIDDHARTA experiment~\cite{Zmeskal:2009zz} at the DA$\Phi$NE electron--positron collider~\cite{Ghigo:2003gy}, and estimates the real part $\Re f_{0}=-0.65\pm0.10$~fm\footnote{The usual femtoscopic sign convention is used, where a positive $\Re f_0$ corresponds to attractive strong interaction.} and the imaginary part $\Im f_{0}=0.81\pm0.15$~fm of the scattering length. Historically, the scattering parameters extracted from the kaonic hydrogen experiments pointed to an attractive interaction, while a repulsive $\kam p$ scattering length in free space was measured via scattering experiments. Currently, both results are explained simultaneously by the presence of the $\Lambda(1405)$ resonance located just below the $\overline{\mathrm{K}}$N threshold~\cite{Hyodo:2011ur}. Similar experiments have been performed with other hadrons, such as pions~\cite{Gotta:2004rq,Hori:2719093} and antiprotons~\cite{Klempt:2002ap}, providing precise measurements of the interaction at the kinematic threshold.

On the other hand, the repulsive nature of the  K$^+$p interaction is well established.
It is described by an s-wave~\cite{Dalitz:1960du} scattering length with a negative real part  ($\Re f_0=-0.308 \pm 0.003$) and the Coulomb interaction~\cite{Hadjimichef:2002xe,Gibbs:2006ab}.

Scattering experiments are special in that both the initial and final states are fixed, which implies that only the $\kam p\rightarrow\kam p$ process can be accessed.
This is not the case for ultrarelativistic particle collisions since only the $\kam p$ final state is fixed and different initial states are allowed, such as $\mathrm{\overline{K}^{0}n}$ and $\uppi\Sigma$, which are referred to as coupled channels. Collisions of nuclei at the Large Hadron Collider (LHC) provide an abundant source of various particle species (including strange hadrons~\cite{ALICE:2017jyt}), in both small (pp, p--Pb) and large (Pb--Pb) systems. In  a single central Pb--Pb collision, the number of produced particles reaches thousands at midrapidity ($|y|<0.5$)~\cite{Abelev:2013xaa,Abelev:2012wca,Abelev:2013vea}. Moreover, since the baryon chemical potential for ultrarelativistic heavy-ion collisions at LHC energies approaches zero, a nearly equal amount of matter and antimatter are produced~\cite{Andronic:2017pug}. Therefore, they are an ideal environment for conducting interaction studies involving both strangeness and antimatter~\cite{Adamczyk:2015hza,STAR:2018uho,Acharya:2019ldv}, as shown in Fig.~\ref{fig:scheme} c. 
A recent breakthrough in the field was achieved by analysing the kaon--proton momentum correlations~\cite{Acharya:2019bsa} using the technique of femtoscopy in high-energy pp collisions at the ALICE experiment~\cite{Aamodt:2008zz}, showing clear evidence for the presence of coupled channels in the measured correlation function. In particular, a clear signature for the opening at the threshold of the $\mathrm{\overline{K}^{0}n}$ isospin breaking channel is observed due to the mass difference between the pairs. 

In order to reproduce the experimental data, theoretical models need to describe the Coulomb interaction, the coupled-channels effects, and the threshold energy difference among the isospin multiplets used in the calculation~\cite{Kamiya:2019uiw}.
This can be done in a realistic $\overline{\rm  K}$N--$\uppi\Sigma$--$\uppi\Lambda$ coupled-channel potential which can be constructed starting from chiral SU(3) dynamics~\cite{Ikeda:2012au}. Alternatively, the $\rm \overline{K}N$ interaction can also be characterized by the Lednick\'y--Lyuboshitz analytical model where the s-wave scattering parameters are used to describe the measurements~\cite{Lednicky:1981su,Lednicky:2003mq}.

It has been recently predicted that the strength of the coupled channels is significantly reduced for source sizes larger than 3~fm, because channels other than $\kam p$ have a considerable magnitude only around a distance of 1~fm~\cite{Kamiya:2019uiw}. It is therefore expected that for larger sources, such as those created in heavy-ion collisions, their effect on the correlation function should be significantly reduced or become asymptotically small.
Therefore, the analysis of $\kam p$ pairs in $\PbPb$ collisions, as shown in this work for the first time, can help to disentangle the $\kam p\rightarrow\kam p$ process from contributions stemming from on-shell coupled channels at the reaction threshold. 

%\section{Methods}
%\label{sec:appendix_methods}

\section{Experimental setup and particle selection}
\label{sec:appendix:ALICE}

The ALICE apparatus~\cite{Aamodt:2008zz} is designed to measure the particles produced in proton and lead (Pb) collisions at the LHC~\cite{Evans:2008zzb}. 
The central barrel consists of a large solenoid magnet used to generate homogeneous magnetic fields up to 0.5 T along the beam directions and a set of detectors located in the interior surrounding the beam axis. 
Additional detectors are located outside the central barrel in the forward beam direction. 
A detailed description of the performance of the ALICE detector can be found in Ref.~\cite{Abelev:2014ffa}.

In this work, the minimum-bias $\PbPb$ data at a collision energy per nucleon--nucleon pair of $\sqrt{s_{\rm NN}}=5.02$ TeV collected by ALICE in the LHC Run 2 (2015) are used. 
The recording of the collisions products was triggered using the V0 detector consisting of two arrays of 32 scintillator counters covering pseudorapidity ($\eta$) ranges of 2.8 $<\eta<$ 5.1 (V0A) and $-3.7<\eta < -1.7$ (V0C).
A coincident signal in both detectors consistent with the collision occurring at the centre of the ALICE detector was required.
The centrality of the collision, expressed in percentages of the total hadronic cross section, was determined using the amplitudes of the signals in the V0 detectors following the procedure described in Ref.~\cite{Abelev:2013qoq} and classified into six intervals of the Pb--Pb cross section: 0--5\%, 5--10\%, 10--20\%, 20--30\%, 30--40\%, and 40--50\%. Those intervals together with the corresponding charged-particle multiplicity densities at midrapidity $\langle \mathrm{d}N_{\rm ch}/\mathrm{d}\eta \rangle$ are listed in Table~\ref{tab:centrality}.
Only collisions within $\pm 10$~cm of the nominal interaction point along the beam axis are considered in this analysis in order to achieve uniform tracking and particle identification performance.

\begin{table}[!h]
\centering
\caption{Centrality ranges and corresponding average charged-particle multiplicity densities at midrapidity $\langle \mathrm{d}N_{\rm ch}/\mathrm{d}\eta \rangle$ for Pb--Pb collisions at $\sqrt{s_{\rm NN}}=5.02$~TeV~\cite{Adam:2015ptt}.}
 \begin{tabular}{  c | c  }

    \hline
    Centrality & $\langle \mathrm{d}N_{\rm ch}/\mathrm{d}\eta \rangle$\\ \hline
    \hline
    0--5\%      & $1943 \pm 53$ \\ \hline
    5--10\%    & $1586 \pm 46$ \\ \hline
    10--20\%  & $1180 \pm 31$ \\ \hline
    20--30\%  & $786 \pm 20$ \\ \hline 
    30--40\%  & $512 \pm 15$ \\ \hline
    40--50\%  & $318 \pm 12$ \\ \hline
\end{tabular}
\label{tab:centrality}
\end{table}

For track reconstruction and hadron identification the analysis uses information provided by the Inner Tracking System (ITS)~\cite{Aamodt:2008zz}, the Time Projection Chamber (TPC)~\cite{Aamodt:2008zz,Dellacasa:2000bm}, and the Time-of-Flight detector (TOF)~\cite{Akindinov:2013tea}, located inside the solenoid magnet.

The ITS is composed of six cylindrical layers of silicon detectors and it is used to determine the locations of the primary vertex by extrapolation of primary tracks. 

The TPC is a large (volume of 88~$\rm m^3$) cylindrical detector filled with gas and is used as the main tracking and particle identification detector of ALICE. The central electrode divides the TPC into two halves. At the end of each half there is a readout plane composed of 18 sectors (with full azimuthal angle $\varphi$ coverage).  Each sector contains 159 padrows arranged radially. A track signal in the TPC consists of space points (referred to as \emph{clusters}), each reconstructed in one of the padrows. A Kalman fit is performed on a set of clusters in order to obtain the parameters of a given track. In this analysis, the determination of momentum is performed using the track curvature from the TPC detector only.

TOF is a cylindrical detector composed of Multigap Resistive Plate Chambers (MRPC) located at $r\cong 380$~cm from the beam axis. They are also arranged into 18 azimuthal sectors. The detector measures the arrival of particles with a precision of the order of 50~ps.

Acceptance is restricted to $|\eta| < 0.8$. 
Selected kaons have transverse momenta of $0.19< p_{\text T} < 0.45$~GeV$/c$, while for protons the interval is $0.5< p_{\text T} < 4.0$~GeV$/c$. 
The distances of closest approach (DCA) of a proton track to the primary vertex in the transverse ($\rm DCA_{\text{XY}}$) and longitudinal ($\rm DCA_{Z}$) directions are required to be less than $0.0105 + 0.0350/(p_{\text{T}}/$(GeV/$c)^{-1.1}$)~cm and 3.2 cm, respectively. 
These selections are imposed to reduce the contamination from  secondary tracks originating from weak decays and from interaction with the detector material.  
All selections are summarized in Table~\ref{tab:trackcuts}.

Particle identification is based on energy loss d$E/\text{d}x$ in the TPC and time of flight in the TOF. The difference between the measured and expected signals can be expressed in units of detector resolution called $N_{\sigma}$. 
For protons, the combined signal from TOF and TPC is allowed to differ from the expected one by $N_{\sigma}=3\sqrt{2}$. For kaons, five selection criteria (depending on the momentum $p$) were introduced, as detailed in Table~\ref{tab:trackcuts}.
The selection criteria are optimised to obtain a high-purity sample without compromising the efficiency.
The identification purity of the data sample is estimated from detailed Monte Carlo (MC) simulations using the HIJING~\cite{Wang:1991hta} event generator coupled to the GEANT3~\cite{Brun:1994aa} transport package and found to be above 98.5\% for both the kaon and proton samples. 
The fraction of protons originating from weak decays is determined to be $5.6\%$ and negligible for kaons ($<0.05\%$).
An additional method based on template fits to the distributions of the ${{\rm DCA}_{\rm Z}}$ is used to determine the contribution from weak decays in the proton sample. 
The fraction of 7.9\% obtained for secondary protons is averaged with the fraction from MC, resulting in a contamination of 6.5\% of weak decays in the sample.

The identified tracks from each event are combined into pairs to form the distribution $A(k^*)$. Two-particle detector acceptance effects, including track splitting, track merging, as well as effects coming from $\rm \gamma\rightarrow e^{+}e^{-}$ conversion, contribute to the measured distributions. 
The following selections are applied to reduce these effects.
For pairs of tracks within the relative pseudorapidity of the two tracks $|\Delta\eta| < 0.01$, an exclusion on the  fraction of merged points is introduced.
The merged fraction is defined as the ratio of the number of steps of $1$ cm considered in the TPC radius range for which the distance between the tracks is less than $3$ cm, to the total number of steps. 
Pairs with a merged fraction above 3\% are removed. The $\rm e^{+}e^{-}$ pairs originating from photon conversions can be misidentified as real kaon--proton pairs and it is necessary to remove spurious correlations arising from such pairs.  
These pairs are removed if their invariant mass, assuming the electron mass for both particles, is less than 0.002 GeV$/c^2$, and  the relative polar angle, $\Delta\theta$, between the two tracks is less than 0.008~rad.

\begin{table}[tb!]
\centering
\caption{Single particle selection criteria for kaons and protons.}
\label{tab:trackcuts}
\begin{tabular}{l|l|l|l}
  \hline
  \multicolumn{4}{c}{Track selection} \\ \hline
   K$^\pm$ $p_{\rm T}$  & \multicolumn{3}{c}{$0.19<p_{\rm T}<0.45$ GeV/$c$} \\ \hline
   p/$\overline{\rm p}$ $p_{\rm T}$  & \multicolumn{3}{c}{$0.5<p_{\rm T}<4$ GeV/$c$} \\ \hline
   $|\eta|$ & \multicolumn{3}{c}{$< 0.8$ }\\ \hline
  $\rm DCA_{\rm z}$ to primary vertex & \multicolumn{3}{c}{$< 3.2$ cm} \\ \hline  
      \hline
  \multicolumn{4}{c}{Kaon selection} \\ \hline
    $N_{\sigma,\rm TPC}$ (for $p < 0.4$~GeV/$c$) & \multicolumn{3}{c}{$< 2$} \\ \hline
    $N_{\sigma,\rm TPC}$ (for $0.4 < p < 0.45$ GeV/$c$) & \multicolumn{3}{c}{$< 1$}  \\ \hline
    $N_{\sigma,\rm TPC}$ (for $p > 0.45$~GeV/$c$) & \multicolumn{3}{c}{$< 3$}  \\ \hline
    $N_{\sigma,\rm TOF}$ (for $0.5 < p < 0.8$ GeV/$c$) & \multicolumn{3}{c}{$< 2$}  \\ \hline
    $N_{\sigma,\rm TOF}$ (for $0.8 < p < 1.0$ GeV/$c$) & \multicolumn{3}{c}{$< 1.5$} \\ \hline
    $N_{\sigma,\rm TOF}$ (for $1.0 < p < 1.5$ GeV/$c$) & \multicolumn{3}{c}{$< 1$} \\ \hline
    
    $\rm DCA_{\rm xy}$ to primary vertex & \multicolumn{3}{c}{2.4~cm} \\ \hline
  \hline
  \multicolumn{4}{c}{Proton selection} \\ \hline
    $\sqrt{N_{\sigma,\rm TPC}^2+N_{\sigma,\rm TOF}^2}$ (for $p_{\rm T} > 0.5$~GeV/$c$) & \multicolumn{3}{c}{$< 3\sqrt{2}$} \\ \hline
    
    $\rm DCA_{\rm xy}$ to primary vertex & \multicolumn{3}{c}{$0.0105 + 0.0350/(p_{\text{T}}/$(GeV/$c)^{-1.1}$)~cm} \\ \hline
\end{tabular}
\end{table}

\newpage
%%%%%%%%%%%%%%%%%%%%%%%%%%%%%%%%%%%%%%%%%%%%%%%%%%%%
%%%%%%% Data analysis %%%%%%%%%%%%%%%%%%%%%%%%%%%%%%%%%
%%%%%%%%%%%%%%%%%%%%%%%%%%%%%%%%%%%%%%%%%%%%%%%%%%%%
\section{Data analysis}

%This analysis uses $\PbPb$ data at a collision energy of $\sqrt{s_{\rm NN}}=5.02$ TeV per nucleon--nucleon collision delivered by the LHC~\cite{Evans:2008zzb} in 2015. Particles produced in the collisions were recorded by ALICE and in total $38$ million collisions were analysed. The description of the experimental setup, details of the (anti-)kaon and (anti-)proton candidate selection, as well as criteria for $\kam p$ ($\kap\overline{p}$) and $\kap p$ ($\kam\overline{p}$) pair formation are discussed in Appendix~\ref{sec:appendix:ALICE}. 

The femtoscopic measurements are based on the sensitivity of two-particle correlations at low relative momentum to the space--time separation
of the particle emitters due to the effects of quantum
statistics and/or final state interactions (FSI), i.e.\ Coulomb and strong forces, depending on the pair under consideration~\cite{Lisa:2005dd,Lednicky:2005tb}. The femtoscopic correlation function, as shown in Fig.~\ref{fig:scheme} c, in the pair rest frame (PRF) can be written as~\cite{Koonin:1977fh,Pratt:1990zq}

\begin{equation}
C(\vec{k}^*) = \int S (\vec{r}^* )~\left | {\Psi  ( \vec{k}^{*},\vec{r}^{*} )} \right |^2 \mathrm{d}^3{r}^*,
\label{eq:cf_def}
\end{equation}

\noindent where $|\vec{k^*}|=\ks$ is half the kaon--proton pair relative momentum, which is equal to the momentum of the first particle in the PRF, where the pair centre of mass is at rest ($\vec{p_1}=-\vec{p_2}$). The two-particle emitting source function is denoted as $S (\vec{r}^*  )$, $\Psi (\vec{k}^{*},\vec{r}^{*}  )$ is the pair wave function, and $\vec{r}^*$ is the relative separation vector. The pair wave function depends on the interactions between the two hadrons~\cite{Lednicky:1981su,Lednicky:2003mq}. 
In the case of equal emission time in the PRF, the Bethe--Salpeter amplitude in the Lednick\'y--Lyuboshitz formalism~\cite{Lednicky:1981su,Lednicky:2003mq,Lednicky:2009zza}, described in detail in Appendix~\ref{sec:appendix:LL},  coincides with a stationary solution of the scattering problem with reversed time direction in the emission process. 
The interaction then depends only on the scattering length, $f_0$, and the effective range  of the interaction, $d_0$. 
In this work, the spin-averaged scattering parameters are obtained, i.e.\ the real and imaginary parts, $\Re f_0$ and $\Im f_0$, respectively, of the scattering length. Moreover, the zero-effective-range approximation ($d_0=0$) is used in the analysis.

The experimental correlation function, $C(\ks)=\EuScript{N}~A(\ks)/B(\ks)$, is constructed as a ratio between the measured distribution of pair relative momenta  $A(\ks)$ and the uncorrelated distribution $B(\ks)$ obtained using the mixed-event technique \cite{Kopylov:1974th,Acharya:2017qtq}, where $\EuScript{N}$ is a free normalization factor which is constrained by $C(\ks)=1$ at  $300<\ks<800$~MeV$/c$. 
Effects of the scattering length as well as of the size of the femtoscopic source are studied in the region of $\ks<150~\mevc$.
Deviations from unity of $C(\ks)$ in this region indicate the presence of an interaction between the studied particles.
In general, the attractive or repulsive forces manifest themselves in values of the correlation function greater or smaller than one, respectively~\cite{Acharya:2020asf}.

The femtoscopic correlation functions for $\kam p$ and $\kap p$ pairs and their charge conjugates $\kap\overline{p}$ and $\kam\overline{p}$ are measured in six centrality intervals~\cite{Abelev:2013qoq}. The centrality intervals and the corresponding charged-particle multiplicity densities at midrapidity $\langle \mathrm{d}N_{\rm ch}/\mathrm{d}\eta \rangle$ are listed in Table~\ref{tab:centrality}. Since no deviations, apart from statistical ones, between pairs and corresponding charge conjugate pairs are observed, they are combined and denoted as $\kap p\oplus \kam\overline{p}$ for same- and $\kam p\oplus \kap\overline{p}$ for opposite-charge combinations, respectively.

In $\PbPb$ collisions, particles are also correlated due to the collective expansion of the system. 
This generates a slope in the correlation function~\cite{Kisiel:2017gip} which 
appears to be  well described by both hydrodynamic simulations~\cite{Bozek:2012qs} coupled to the statistical hadronisation code THERMINATOR 2~\cite{Chojnacki:2011hb} or using the AMPT model~\cite{Lin:2004en}. 
These background correlations are therefore subtracted before fitting the correlation functions by employing a first-order polynomial fit to the AMPT simulated data in the first two centrality intervals (0--10\%) and a fit to the THERMINATOR 2 simulated data for the remaining centrality classes (10--50\%). The fixed value of the impact parameter used for 0--10\% centrality interval in THERMINATOR 2 was not reproducing the distribution of measured average charged-particle multiplicity.  

%%%%%%%%%%%%%%%%%%%%%%%%%%%%%%%%%%%%%%%%%%%%%%%%%%%%
%%%%%%% fits %%%%%%%%%%%%%%%%%%%%%%%%%%%%%%%%%%%%%%%
%%%%%%%%%%%%%%%%%%%%%%%%%%%%%%%%%%%%%%%%%%%%%%%%%%%%

The extraction of the scattering length parameters of $\kam p\oplus\kap\overline{p}$ pairs from the measured correlation functions uses a dedicated fitting procedure based on the Lednick\'y--Lyuboshitz formalism. 
A simultaneous fit to all measured pairs in the six centrality classes is performed in the range up to $\ks=175~\mevc$. The final values of the fitted parameters are obtained from the comparison of the experimental correlation functions to the set of precomputed model functions and based on the lowest $\chi^2$/ndf.

The assumption for the source function $S(\vec{r}^*)$ is that, for each centrality, it is a three-dimensional spheroid with a Gaussian density profile in the PRF~\cite{Broniowski:2008vp,Acharya:2017qtq,Acharya:2020nyr} with a common width parameter for same- and opposite-charge pairs $R_{\rm Kp}$, referred to as the source size or femtoscopic radius. This assumption is based on hydrodynamic models~\cite{Huovinen:2006jp,Gale:2013da} of the matter produced in ultrarelativistic heavy-ion collisions which provide quantitative agreement~\cite{Foka:2016vta} with collective phenomena~\cite{Adam:2016izf}, particle transverse-momentum spectra~\cite{Abelev:2014laa}, and femtoscopy of both identical~\cite{Adam:2015vja,Acharya:2017qtq} and non-identical particles~\cite{Acharya:2020nyr}. 
The femtoscopic radii follow the previously observed linear scaling with $\langle\mathrm{d}N_{\rm ch}/\mathrm{d}\eta\rangle^{1/3}$ and the transverse-mass scaling behaviour~\cite{Adamova:2002wi,Adams:2004yc,Abelev:2009tp,Adam:2015vna,Adam:2015vja}. 
This allows one to fix the source functional form in Eq.~(\ref{eq:cf_def}) and use the Lednick\'y--Lyuboshitz formalism for measurements of the interaction between any pair of particles that can be detected in the final state.
Moreover, the fit is further constrained by using the measured scattering length values for $\kap p\oplus\kam\overline{p}$ pairs, established from a partial wave analysis to be $\Re f_0=-0.308 \pm 0.003$~fm~\cite{Gibbs:2006ab}. 
Finally, there are six different femtoscopic radii corresponding to each centrality class and two, real and imaginary, components of the $\kam p\oplus\kap\overline{p}$ scattering length. 
In a simultaneous fit to all measured correlation functions, the fitting procedure determines the six radii from the $\kap p$ correlations and the real and imaginary parameters of the strong interaction scattering length from the $\kam p$ correlations.

The fitting procedure also accounts for the purity of the sample, defined as the percentage of properly
identified primary particle pairs originating from the three-dimensional Gaussian profile of the source, following the method described in Ref.~\cite{Kisiel:2009eh}. This method was  successfully employed in the pion--kaon femtoscopic analysis by ALICE~\cite{Acharya:2020nyr}. The value for the purity parameter was estimated to be 87\% and depends on the misidentification, secondary contamination from weak decays, and percentage of kaons and protons that come from strongly decaying resonances constituting the long tails in the source distribution, outside the Gaussian core. 

The measured correlation function is also affected by the finite momentum resolution of the ALICE detector. The detector response matrix is obtained using a detailed Monte Carlo simulation of the detector based on HIJING~\cite{Wang:1991hta} coupled to GEANT3~\cite{Brun:1994aa} transport code and is used to smear the theoretical calculations. 

\section{Systematic uncertainties}
\label{app:syst_unc}
%%%%%%%%%%%%%%%%%%%%%%%%%%%%%%%%%%%%%%%%%%%%%%%%%%%%
%%%%%%% SYSTEMATIC ERRORS %%%%%%%%%%%%%%%%%%%%%%%%%%
%%%%%%%%%%%%%%%%%%%%%%%%%%%%%%%%%%%%%%%%%%%%%%%%%%%%

The final systematic uncertainties are the maximum and minimum deviations from the reference values obtained for different variations of the analysis. 
The relative values of the important sources of uncorrelated systematic uncertainty are listed in Table~\ref{table:1}. 
One of the most significant sources of uncertainty comes from  background estimation. 
The analysis considers two variations of the default option. The first one (Background 1) defines the background as a first-order polynomial fit to the raw data in the $k^{*}$ region 0.35--0.6 GeV/$c$ instead of predictions from the AMPT model in the first two centrality intervals. 
The second one (Background 2) estimates the background from the average of the first-order polynomial fit to the raw data and the AMPT prediction in the first two centrality intervals and the average of functions provided by the THERMINATOR 2 code and the AMPT model in the other four centrality intervals.
The pair purity value used for scaling data points is also varied. 
The default value equals  87\%, the effect of possible additional sources of impurities is evaluated and found to be smaller than 7\%. 
The systematic effect of the finite resolution of the detectors is evaluated using an alternative technique based on a smoothed detector response matrix.
Another variation of the default analysis is obtained by using a different sample of kaons. 
The default data contain kaons in the transverse-momentum range between 0.19--0.45 GeV/$c$, while the variations cover the regions of 0.5--1.5 GeV/$c$ as well as 0.19--1.5 GeV/$c$. The sample with high $p_{\rm T}$ is the most significant source of the uncertainty of the real and imaginary parts of the scattering length due to the change in the average pair transverse mass $m_{\rm T}$ (the changes of radii are expected and should not be treated as systematic uncertainty). The systematic uncertainty related to the DCA is evaluated using an alternative selection of a $p_{\rm T}$ dependent ${\rm DCA}_{\rm Z}<0.418-0.372\times p_{\rm T}^{0.66}$~cm and a constant value of $\rm DCA_{\rm XY}<2.4$~cm, preserving the integrated proton purity. 
Additionally, results from the combined fit to both kaon selections are also taken into account where the high $p_{\rm T}$ kaon sample radii are scaled to the low-$p_{\rm T}$ according to the average pair $m_{\rm T}$. 
Variations of the fitting range and track reconstruction have a negligible contribution compared with other sources. 

\begin{table} [tbh!]
\caption{Components of the systematic uncertainty expressed as relative uncertainties from various sources. For a detailed description of each variation see the text. }
\centering
\fontsize{8}{5}\small\addtolength{\tabcolsep}{-0.1pt}
\begin{tabular}{ l | c  c  c  c  c  c | c | c  }
\hline
 	&	\multicolumn{6}{c|}{$\Delta R_{\rm Kp}$ (\%)} &		&		\\ \cline{2-7}
Variation 	&\multicolumn{6}{c|}{Centrality}	&		$\Delta \Re f_0$ (\%)	&		$\Delta \Im f_0$ (\%)	\\ 
 	&		0--5\%	&	5--10\% &	10--20\%	&	20--30\% &	30--40\%	&	40--50\%	&	&	\\ \hline \hline
Background 1 	&	4	&	$<$1	&	$<$1	&	$<$1	&	$<$1	&	$<$1	&	3	&	1	\\ \hline
Background 2	&	3	&	3	&	4	&	3	&	1	&	3	&	6	&	13	\\ \hline
Pair purity $-3$\%	&	$-$3	&	$-$4	&	$-$3	&	$-$3	&	$-$3	&	$-$3	&	$+$1	&	$-$5	\\ \hline
Pair purity +3\%	&	$+$6	&	$+$4	&	$+$7	&	$+$4	&	$+$4	&	$+$4	&	$-$2	&	$+$11	\\ \hline
Momentum 	&	\multirow{2}{*}{$-$14}	&	\multirow{2}{*}{$-$13}	&	\multirow{2}{*}{$-$9}	&	\multirow{2}{*}{$-$11}	&	\multirow{2}{*}{$-$10}	&	\multirow{2}{*}{$-$10}	&	\multirow{2}{*}{$+$19}	&	\multirow{2}{*}{$-$27}	\\
resolution 	&		&		&		&		&		&		&\\
 \hline

Fit range	&	\multirow{2}{*}{5}	& \multirow{2}{*}{2}	&	\multirow{2}{*}{1}	&	\multirow{2}{*}{1}	&	\multirow{2}{*}{$<$1}	&	\multirow{2}{*}{$<$1}	& \multirow{2}{*}{2}	&	\multirow{2}{*}{$<$1}	\\ 
0--0.2 GeV$/c$ &		&		&		&		&		&		&\\\hline

Fit range	&	\multirow{2}{*}{3}	& \multirow{2}{*}{3}	&	\multirow{2}{*}{8}	&	\multirow{2}{*}{$<$1}	&	\multirow{2}{*}{$<$1}	&	\multirow{2}{*}{$<$1}	& \multirow{2}{*}{4}	&	\multirow{2}{*}{1}	\\ 
0.005--0.150 GeV$/c$ &		&		&		&		&		&		&\\\hline

Alternative DCA  	& \multirow{2}{*}{8}	&	\multirow{2}{*}{6}	&	\multirow{2}{*}{4}	&	\multirow{2}{*}{6}	&	\multirow{2}{*}{$<$1}	&	\multirow{2}{*}{7} 	&	\multirow{2}{*}{1}	&	\multirow{2}{*}{5} \\ selection 	&		&		&		&		&		&		&\\ \hline 
Kaon selection &	\multirow{2}{*}{$-$11}	&	\multirow{2}{*}{$-$8}	&	\multirow{2}{*}{$-$8}	&	\multirow{2}{*}{$-$13}	&	\multirow{2}{*}{$-$5}	&	\multirow{2}{*}{$-$4}	&	\multirow{2}{*}{$+$16}	&	\multirow{2}{*}{$-$10}	\\ 
(low and high $p_{\rm T}$)		&		&		&		&		&		&		& \\ \hline
Kaon selection  	&	\multicolumn{6}{c|}{\multirow{2}{*}{Does not apply}}		&	\multirow{2}{*}{$+$19}	&	\multirow{2}{*}{$-$35}	\\ 
(high $p_{\rm T}$ only) 	&		&		&		&		&		&		&	\\ \hline
\end{tabular}
\label{table:1}
\end{table} 
%%%%%%%%%%%%%%%%%%%%%%%%%%%%%%%%%%%%%%%%%%%%%%%%%%%%
%%%%%%% RESULTS  %%%%%%%%%%%%%%%%%%%%%%%%%%%%%%%%%%%
%%%%%%%%%%%%%%%%%%%%%%%%%%%%%%%%%%%%%%%%%%%%%%%%%%%%
\section{Results}

\begin{figure}[tb!]
    \centering
    \includegraphics[width=1.0\textwidth]{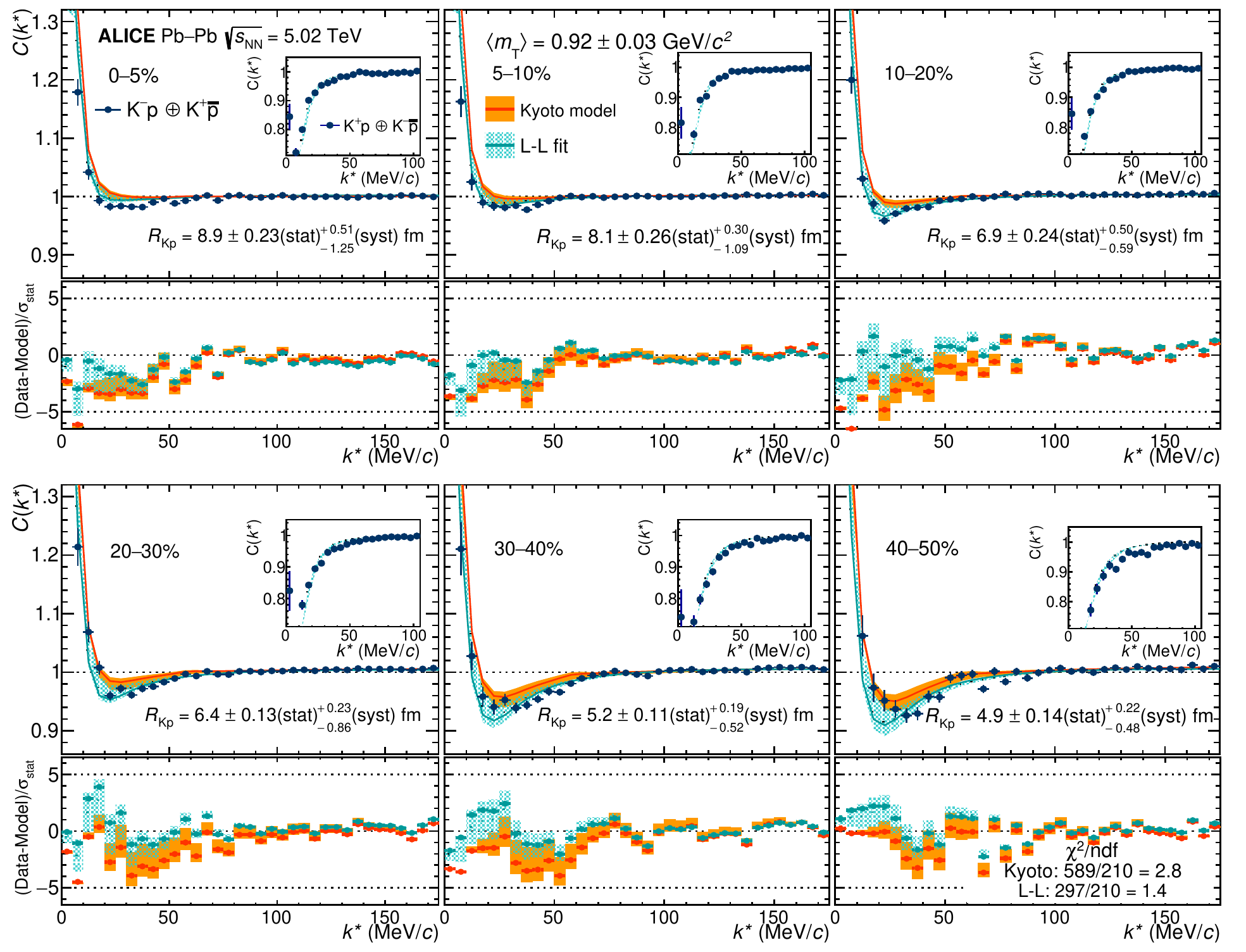}
    \caption{
    The $\rm K^{-}p \oplus K^{+}\overline{p}$ correlation functions in the six centrality classes, 
     with the corresponding  Lednick\'y--Lyuboshitz fits (denoted as ``L--L'') and Kyoto model calculations shown as light cyan and orange bands, respectively. The  width of the bands corresponds to the 1-$\sigma$ uncertainties. The inserts show the $\rm K^{+}p \oplus K^{-}\overline{p}$ correlation functions with Lednick\'y--Lyuboshitz fits as light cyan bands. The bottom panels show the difference between data and the fit (model) normalised by the statistical uncertainty of the data $\sigma_{\rm stat}$. 
     The average pair transverse mass $\langle m_{\rm T}\rangle$ is $0.92\pm0.03$~GeV/$c^{2}$ for all centrality intervals.
    The statistical and systematic uncertainties are added in quadrature and shown as vertical bars.
    }
    \label{fig:cfs}
\end{figure}

\begin{figure}[tb]
 \begin{minipage}{0.495\textwidth}
     \centering
     \includegraphics[width=0.99\textwidth]{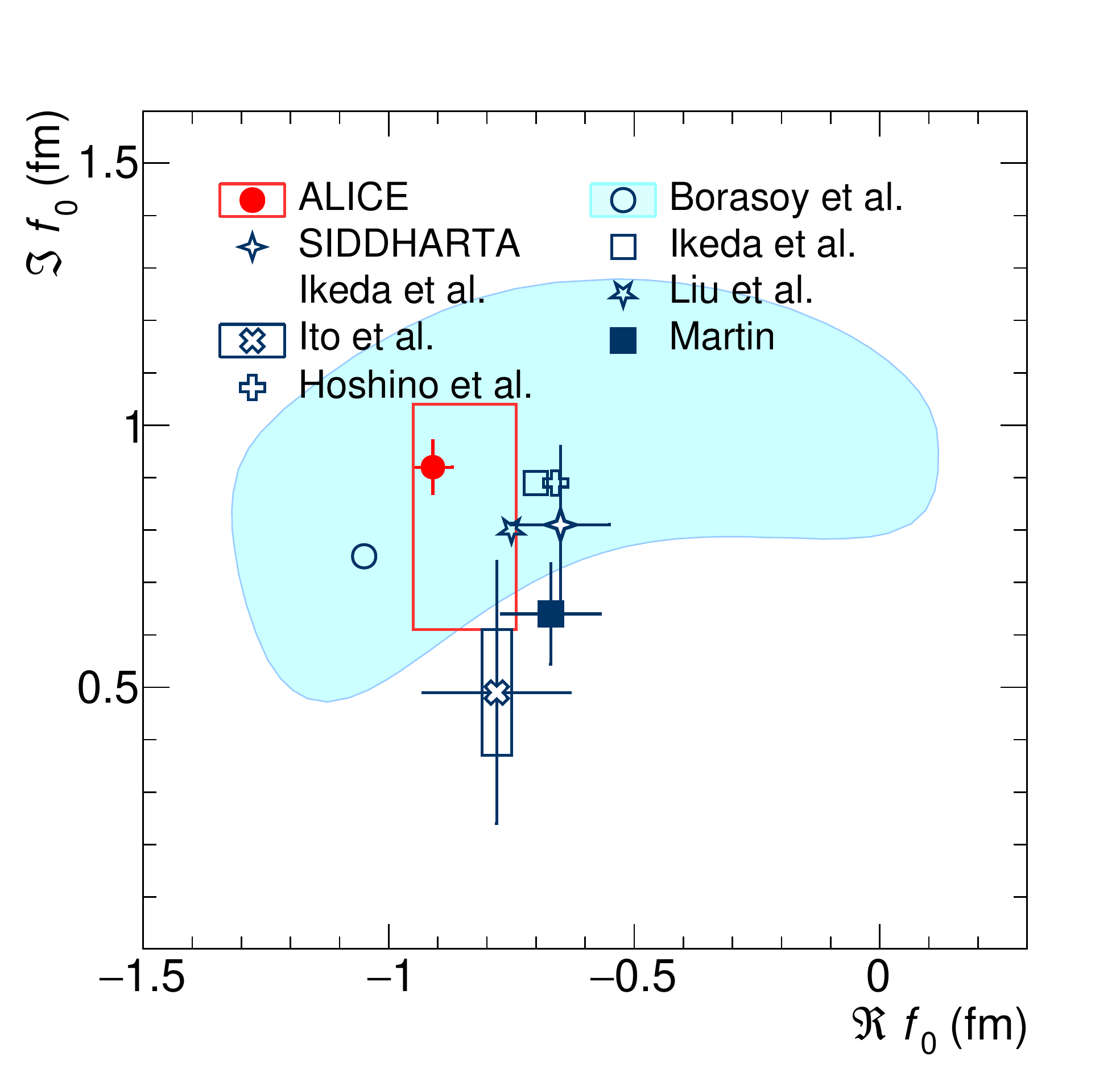}
     \end{minipage}\hfill
  \begin{minipage}{0.495\textwidth}
    \centering
     \includegraphics[width=0.99\textwidth]{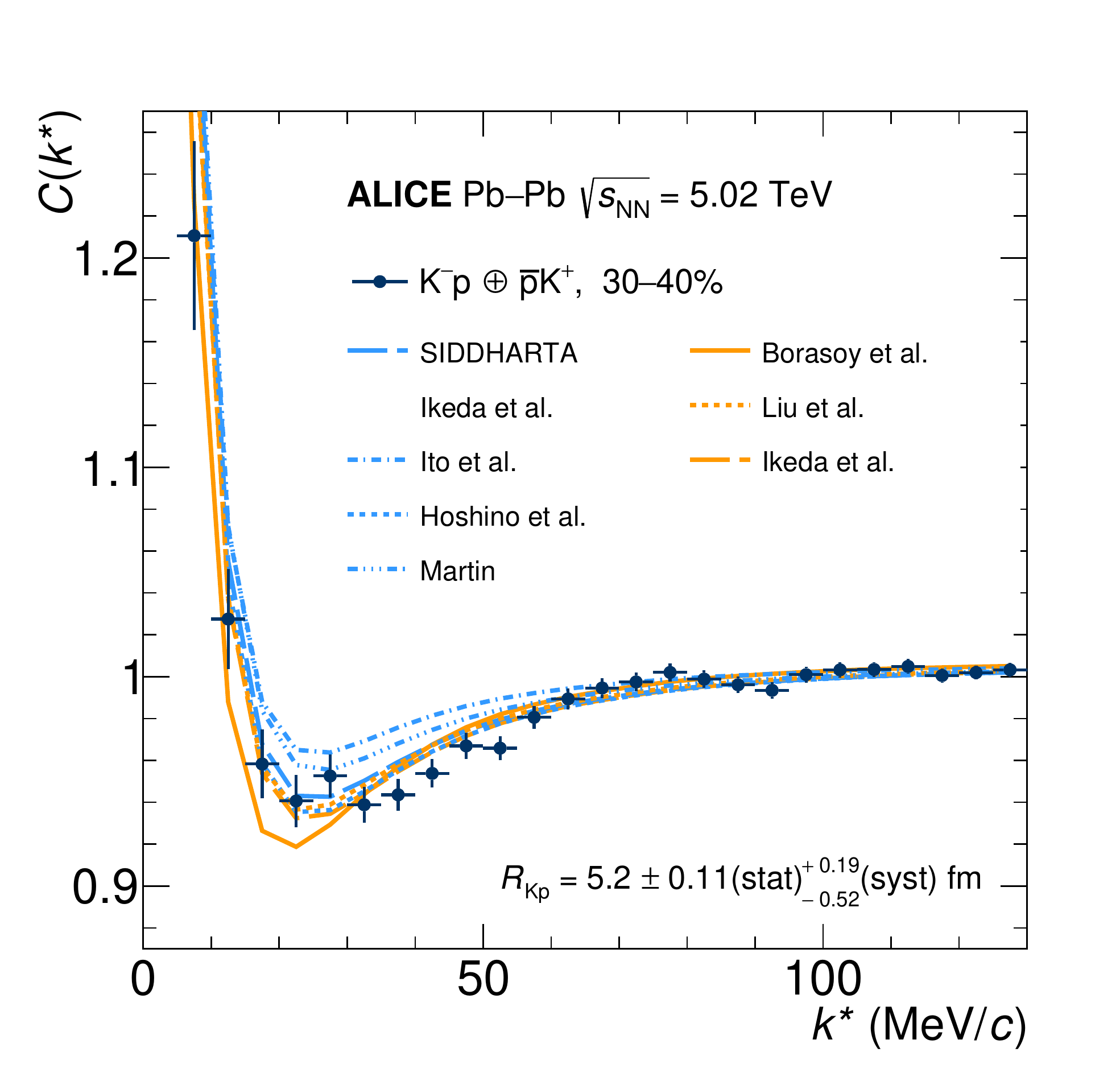} 
     \end{minipage}
        \caption{Left: scattering parameters obtained from the Lednick\'y--Lyuboshitz  fit compared with available world data and theoretical calculations.  Statistical uncertainties are represented as bars and systematic uncertainties, if provided, as boxes. Right: experimental femtoscopic correlation function for $\kam p\oplus \kap \overline{p}$ pairs in the 30--40\% centrality interval, together with various Lednick\'y--Lyuboshitz calculations obtained using the scattering length parameters from Refs.~\cite{Martin:1980qe,Ito:1998yi,Liu:2020foc,Ikeda:2011pi,Ikeda:2012au,Hoshino:2017mty,Borasoy:2006sr} and the source radius from this analysis. The statistical and systematic uncertainties of the measured data points are added in quadrature and shown as vertical bars.}
            \label{fig:scatt_params}
\end{figure}

The correlation functions for opposite-charge particle pairs ($\kam\p\oplus\kap\overline{p}$), after division by the AMPT or THERMINATOR 2 baselines, are shown in Fig.~\ref{fig:cfs}. The correlation functions for same-charge particle pairs ($\kap\p\oplus\kam\overline{p}$) are presented as inserts. %Details of the systematic uncertainty estimation are discussed in Appendix~\ref{app:syst_unc}.
The cyan band represents the result of the simultaneous fit to all obtained correlation functions with the Lednick\'y--Lyuboshitz  model~\cite{Lednicky:1981su,Lednicky:2003mq}. 
 The orange band corresponds to predictions from $\rm \chi^{EFT}$ calculations. These are computed using the Kyoto model~\cite{Kamiya:2019uiw} and evaluated with the CATS framework~\cite{Mihaylov:2018rva}. 
The Kyoto model is based on a chiral effective SU(3) calculation in which all coupled channels, such as  $\rm \overline{K}^0$n and $\uppi\Sigma$ are included. At present, this model is able to describe both the ALICE results from proton--proton collisions~\cite{Acharya:2019bsa} and the SIDDHARTA measurement~\cite{Ikeda:2012au}. The bottom panels of Fig.~\ref{fig:cfs} show the difference between the data and the fit (model) divided by the statistical uncertainty of the data, $\rm \sigma_{\rm stat}$.

The following effects can be observed: the $\kam\p\oplus\kap\overline{p}$ pairs show an attractive Coulomb interaction for small $k^*$. The effect is opposite for $\kap\p\oplus\kam\overline{p}$ pairs. The influence of the repulsive strong interaction manifests as correlation functions reaching values below unity in the region of $\ks\approx20-50$~MeV$/c$  and becomes more pronounced towards more peripheral events, i.e., smaller source sizes. 
As predicted in Ref.~\cite{Kamiya:2019uiw}, features of the correlation function related to the coupled channels, observed in the analysis of pp collisions~\cite{Acharya:2019bsa}, are negligible here. Neither the cusp structure at 58 MeV/$c$ due to the presence of the isospin-breaking channel ${\rm K}^0\n\rightarrow \kam\p$ nor the enhancement due to the coupled channels below threshold enhancing the correlation above unity in the intermediate $k^*$ range are visible in the correlation function in Pb--Pb.  

The common femtoscopic radii $R_{\rm Kp}$ for same- and opposite-charge pairs obtained from the Lednick\'y--Lyuboshitz fit are provided in Fig.~\ref{fig:cfs} as well. 
They increase from around 5~fm for peripheral events to almost 9~fm for central events, and all are larger than 3~fm where the predicted effect of coupled channels is reduced or negligible~\cite{Kamiya:2019uiw}. The radii scale linearly with the cube root of the mean charged-particle multiplicity density $\langle \mathrm{d}N_{\rm ch}/\mathrm{d}\eta \rangle^{1/3}$, as observed for pion--pion~\cite{Adam:2015vna}, kaon--kaon~\cite{Acharya:2017qtq}, and pion--kaon~\cite{Acharya:2020nyr} pairs. 
The scattering length parameters obtained from the fit are $\Re f_0=-0.91\pm~0.03$(stat)$^{+0.17}_{-0.03}$(syst)~fm and $\Im f_0 = 0.92\pm~0.05$(stat)$^{+0.12}_{-0.33}$(syst)~fm. 

The obtained parameters of the scattering length are compared with the available experimental values as well as model calculations~\cite{Martin:1980qe,Ito:1998yi,Liu:2020foc,Ikeda:2012au, Hoshino:2017mty,Borasoy:2006sr} in the left panel of Fig.~\ref{fig:scatt_params}. Numerical values of those parameters are also provided in Tab.~\ref{tab:chi2_all_models}. The ALICE results are compatible with them within uncertainties\footnote{Note that systematic uncertainties are not provided for some of the older results.}. Up until this point, the world's best experimental data on Kp scattering are mainly from exotic kaonic atoms, where the interaction at the threshold is measured, and from scattering experiments. Theory predictions and calculations are based on $\rm \chi^{EFT}$ models.

Moreover, the Lednick\'y--Lyuboshitz formalism is also used to compute femtoscopic correlation functions using scattering length parameters from previous measurements and theory predictions. They are then compared with the experimental data and the deviations in units of $\chi^2/\mathrm{ndf}$ are obtained. The result of such a procedure is shown in Fig.~\ref{fig:scatt_params}~(right), while the $\chi^2/\mathrm{ndf}$ values are presented in Table~\ref{tab:chi2_all_models}. The Kyoto model, which captures well the structures related to coupled channels in pp collisions, reproduces the data trends in all measured Pb--Pb centrality intervals, confirming that the coupled channels are fundamental in the description of small sources but have a negligible influence on correlation functions at large source sizes~\cite{Kamiya:2019uiw}. However, the model still requires further development as the resulting $\chi^2/\mathrm{ndf}=2.8$ is slightly worse than the best calculations using the Lednick\'y--Lyuboshitz analytical approach.

\begin{table}[htb!]
	\centering
	\caption{Values of the scattering parameters and the $\chi^2/\mathrm{ndf}$ for the deviation between the ALICE data and available model calculations and previous measurements for $\kam$p pairs at low relative momentum.}
	\label{tab:chi2_all_models}
	{\fontsize{8}{5}\selectfont\small\addtolength{\tabcolsep}{-5pt}
		\begin{tabular}{ l | c | c | c }
			\hline
			\hline
			{\bf Model calculation:} & $\Re f_0$ (fm) & $\Im f_0$ (fm) &   $\chi^2/{\text{ndf}}$  \\
			\hline
			\hline
			Lednick\'y--Lyuboshitz fit to data & $\ -$0.91{$\pm~0.03$(stat)}$^{+0.17}_{-0.03}$(syst)$\ $  & $\ $0.92$\pm~0.05$(stat)$^{+0.12}_{-0.33}$(syst)$\ $  & 1.4 \\
			\hline
			Kyoto \cite{Kamiya:2019uiw,Mihaylov:2018rva}& -- & --& 2.8\\
			\hline
			\hline
			\multicolumn{4}{l}{
				{\bf Lednick\'y--Lyuboshitz with fixed parameters from:}} \\
			\hline
			\hline
			Kaonic deuterium (Hoshino et al.) \cite{Hoshino:2017mty}& $-$0.66 & 0.89 & 2.0\\
			Scattering experiments  (Martin)  \cite{Martin:1980qe}& $-$0.67$\pm$0.1 & 0.64$\pm$0.1  & 3.3\\
			Chiral SU(3) (Ikeda et al.) \cite{Ikeda:2011pi,Ikeda:2012au} & $-$0.7 & 0.89 & 1.9\\
			SIDDHARTA chiral SU(3) \cite{Ikeda:2011pi,Ikeda:2012au} & $-$0.65$\pm$0.1 & 0.81$\pm$0.15 & 2.3\\
			Hamiltonian EFT (Liu et al.) \cite{Liu:2020foc}& $-$0.75 & 0.80  & 1.9\\
			Kaonic hydrogen (Ito et al.)  \cite{Ito:1998yi}& $-$0.78$\pm$0.15 & 0.49$\pm$0.25  & 4.2\\
			Chiral SU(3)  (Borasoy et al.) \cite{Borasoy:2006sr} & $-$1.05$\pm$0.5 & 0.75$\pm$0.4 & 1.6\\
			\hline
	\end{tabular}}
\end{table}

\newpage

\section{Summary}
In this work, we present the first measurement of non-identical particle femtoscopy of $\kam p\oplus \kap\overline{p}$ and $\kap p\oplus \kam\overline{p}$ pairs performed in  Pb--Pb collisions at $\sqrt{s_{\text{NN}}}~=$~5.02 TeV in six centrality intervals. 
Up to now, the existing experimental data on $\kam p$ scattering allowed us to study the interaction at a fixed range, either in the asymptotic regime, where the effect of the coupled channels is not present, or at very small ranges ($R_\mathrm{Kp}\sim 1$~fm) where they are dominant and difficult to disentangle from the $\kam p\rightarrow\kam p$ channel. The analysis of the Pb--Pb data with the ALICE detector addresses these issues by testing the $\kam p$ interaction as a function of the source size in the range $R_\mathrm{Kp}\in(4.9, 8.9)$~fm. In femtoscopy, a precise knowledge of the source from which hadrons are emitted is needed to measure the interaction using quantum scattering theory. In the case of pp collisions, where this is of the order of 1~fm, very short interaction distances are probed. 
In Pb--Pb collisions, the larger average distance of several fm between the hadrons  probes the asymptotic form of the two-particle wave function, as demonstrated in this work. The obtained values of $R_{\rm Kp}$ follow a linear scaling with the cube root of averaged charged-particle multiplicity density and the transverse-mass scaling. This is in agreement with other femtoscopic measurements in heavy-ion collisions, as expected from the hydrodynamic models.  The resulting correlation functions are fitted simultaneously to extract the complex scattering length whose components are $\Re f_0=-0.91\pm0.03$(stat)$^{+0.17}_{-0.03}$(syst)~fm and $\Im f_0 = 0.92\pm0.05$(stat)$^{+0.12}_{-0.33}$(syst)~fm. The obtained result is consistent with existing data from scattering experiments. Moreover, theoretical computations based on $\chi^{\rm EFT}$ (the Kyoto model) provide a comparably good description across all centrality intervals, providing further confirmation on the validity of the model.

%%%%%%%%%%%%%%%%%%%%%%%%%%%%%%%%
% end main text 
%%%%%%%%%%%%%%%%%%%%%%%%%%%%%%%%

%%%%% acknowledgements - handled by EB chairs 
\newenvironment{acknowledgement}{\relax}{\relax}
\begin{acknowledgement}
\section*{Acknowledgements}
% add specific acknowledgements here 
% ...but please don't remove the line below: funding agencies
% will be acknowledged with a custom tex file handled by EB chairs after Collab Round 2
% Version: 2021-04-28

The ALICE Collaboration would like to thank all its engineers and technicians for their invaluable contributions to the construction of the experiment and the CERN accelerator teams for the outstanding performance of the LHC complex.
The ALICE Collaboration gratefully acknowledges the resources and support provided by all Grid centres and the Worldwide LHC Computing Grid (WLCG) collaboration.
The ALICE Collaboration acknowledges the following funding agencies for their support in building and running the ALICE detector:
A. I. Alikhanyan National Science Laboratory (Yerevan Physics Institute) Foundation (ANSL), State Committee of Science and World Federation of Scientists (WFS), Armenia;
Austrian Academy of Sciences, Austrian Science Fund (FWF): [M 2467-N36] and Nationalstiftung f\"{u}r Forschung, Technologie und Entwicklung, Austria;
Ministry of Communications and High Technologies, National Nuclear Research Center, Azerbaijan;
Conselho Nacional de Desenvolvimento Cient\'{\i}fico e Tecnol\'{o}gico (CNPq), Financiadora de Estudos e Projetos (Finep), Funda\c{c}\~{a}o de Amparo \`{a} Pesquisa do Estado de S\~{a}o Paulo (FAPESP) and Universidade Federal do Rio Grande do Sul (UFRGS), Brazil;
Ministry of Education of China (MOEC) , Ministry of Science \& Technology of China (MSTC) and National Natural Science Foundation of China (NSFC), China;
Ministry of Science and Education and Croatian Science Foundation, Croatia;
Centro de Aplicaciones Tecnol\'{o}gicas y Desarrollo Nuclear (CEADEN), Cubaenerg\'{\i}a, Cuba;
Ministry of Education, Youth and Sports of the Czech Republic, Czech Republic;
The Danish Council for Independent Research | Natural Sciences, the VILLUM FONDEN and Danish National Research Foundation (DNRF), Denmark;
Helsinki Institute of Physics (HIP), Finland;
Commissariat \`{a} l'Energie Atomique (CEA) and Institut National de Physique Nucl\'{e}aire et de Physique des Particules (IN2P3) and Centre National de la Recherche Scientifique (CNRS), France;
Bundesministerium f\"{u}r Bildung und Forschung (BMBF) and GSI Helmholtzzentrum f\"{u}r Schwerionenforschung GmbH, Germany;
General Secretariat for Research and Technology, Ministry of Education, Research and Religions, Greece;
National Research, Development and Innovation Office, Hungary;
Department of Atomic Energy Government of India (DAE), Department of Science and Technology, Government of India (DST), University Grants Commission, Government of India (UGC) and Council of Scientific and Industrial Research (CSIR), India;
Indonesian Institute of Science, Indonesia;
Istituto Nazionale di Fisica Nucleare (INFN), Italy;
Institute for Innovative Science and Technology , Nagasaki Institute of Applied Science (IIST), Japanese Ministry of Education, Culture, Sports, Science and Technology (MEXT) and Japan Society for the Promotion of Science (JSPS) KAKENHI, Japan;
Consejo Nacional de Ciencia (CONACYT) y Tecnolog\'{i}a, through Fondo de Cooperaci\'{o}n Internacional en Ciencia y Tecnolog\'{i}a (FONCICYT) and Direcci\'{o}n General de Asuntos del Personal Academico (DGAPA), Mexico;
Nederlandse Organisatie voor Wetenschappelijk Onderzoek (NWO), Netherlands;
The Research Council of Norway, Norway;
Commission on Science and Technology for Sustainable Development in the South (COMSATS), Pakistan;
Pontificia Universidad Cat\'{o}lica del Per\'{u}, Peru;
Ministry of Education and Science, National Science Centre and WUT ID-UB, Poland;
Korea Institute of Science and Technology Information and National Research Foundation of Korea (NRF), Republic of Korea;
Ministry of Education and Scientific Research, Institute of Atomic Physics and Ministry of Research and Innovation and Institute of Atomic Physics, Romania;
Joint Institute for Nuclear Research (JINR), Ministry of Education and Science of the Russian Federation, National Research Centre Kurchatov Institute, Russian Science Foundation and Russian Foundation for Basic Research, Russia;
Ministry of Education, Science, Research and Sport of the Slovak Republic, Slovakia;
National Research Foundation of South Africa, South Africa;
Swedish Research Council (VR) and Knut \& Alice Wallenberg Foundation (KAW), Sweden;
European Organization for Nuclear Research, Switzerland;
Suranaree University of Technology (SUT), National Science and Technology Development Agency (NSDTA) and Office of the Higher Education Commission under NRU project of Thailand, Thailand;
Turkish Energy, Nuclear and Mineral Research Agency (TENMAK), Turkey;
National Academy of  Sciences of Ukraine, Ukraine;
Science and Technology Facilities Council (STFC), United Kingdom;
National Science Foundation of the United States of America (NSF) and United States Department of Energy, Office of Nuclear Physics (DOE NP), United States of America.
\end{acknowledgement}

%%%%%%%% Bibliography 

\bibliographystyle{utphys} % Remember we use title in the biblio
\bibliography{bibliography}
%\input {bibliography.tex}  

%%%%%%%%%%%%%%%%%%%%%%%%%%%%%%%%
% Appendices: yours (if any) + authorlist
%%%%%%%%%%%%%%%%%%%%%%%%%%%%%%%%
\newpage
\appendix

%%%%% Authorlist - please do not touch: handled by EB chairs 
\section{The ALICE Collaboration}
\label{app:collab}

\begin{flushleft}

S.~Acharya$^{\rm 143}$, 
D.~Adamov\'{a}$^{\rm 98}$, 
A.~Adler$^{\rm 76}$, 
J.~Adolfsson$^{\rm 83}$, 
G.~Aglieri Rinella$^{\rm 35}$, 
M.~Agnello$^{\rm 31}$, 
N.~Agrawal$^{\rm 55}$, 
Z.~Ahammed$^{\rm 143}$, 
S.~Ahmad$^{\rm 16}$, 
S.U.~Ahn$^{\rm 78}$
% ...

\bigskip

\bigskip 

\textbf{\Large Affiliation Notes}

\bigskip 

% ...
$^{\rm V}$ Also at: Institute of Theoretical Physics, University of Wroclaw, Poland\\

\bigskip

\bigskip 

\textbf{\Large Collaboration Institutes}

\bigskip 

% ...
$^{16}$ Department of Physics, Aligarh Muslim University, Aligarh, India\\
% ...
$^{31}$ Dipartimento DISAT del Politecnico and Sezione INFN, Turin, Italy\\
% ...
$^{35}$ European Organization for Nuclear Research (CERN), Geneva, Switzerland\\
% ...
$^{55}$ INFN, Sezione di Bologna, Bologna, Italy\\
% ...
$^{76}$ Johann-Wolfgang-Goethe Universit\"{a}t Frankfurt Institut f\"{u}r Informatik, Fachbereich Informatik und Mathematik, Frankfurt, Germany\\
% ...
$^{78}$ Korea Institute of Science and Technology Information, Daejeon, Republic of Korea\\
% ...
$^{83}$ Lund University Department of Physics, Division of Particle Physics, Lund, Sweden\\
% ...
$^{98}$ Nuclear Physics Institute of the Czech Academy of Sciences, \v{R}e\v{z} u Prahy, Czech Republic\\
% ...
$^{143}$ Variable Energy Cyclotron Centre, Homi Bhabha National Institute, Kolkata, India\\
% ...

\end{flushleft}

\subsection{Lednick\'y--Lyuboshitz formalism}
\label{sec:appendix:LL}

The pair wave function in Eq.~(\ref{eq:cf_def}), $\Psi  ( \vec{k}^{*},\vec{r}^{*} )$, depends on the two-particle interaction. Kaons  interact with protons via the strong and, for charged kaons, also the Coulomb force. In such a scenario, the interaction of two non-identical particles is given by the Bethe--Salpeter amplitude, corresponding to the solution of the quantum scattering problem taken in the inverse time direction\footnote{The inverse time direction is manifested by the ``$-$'' sign in front of $k^*$.}:

\begin{equation}
\Psi^{(+)}_{-{\vec{k}^{*}}}({\vec{{r}^{*}}}) = \sqrt{A_{\rm C} (\varepsilon)}
\frac{1}{\sqrt{2}} \left [ {\rm e}^{-{\rm i} \vec{ k}^{*}\times{\vec{r}^{*}}} {\rm F}(-{\rm i} \varepsilon, 1,
  {\rm i} \zeta^{+}) + f_{\rm C}(\vec{k}^*)\frac{\tilde{G}(\rho,\varepsilon)}{{r}^*} \right ],
\label{eq:fullpsi2}
\end{equation}
where $A_{\rm C}$ is the Gamow factor, $\zeta^{\pm} = k^{*} r^{*} (1 \pm \cos{\theta^{*}})$, $\varepsilon = 1/(k^{*} a_{\rm C})$, $\rm F$ is the confluent hypergeometric function, and $\tilde{G}$ is the combination of the regular and singular s-wave Coulomb functions. The angle $\theta^{*}$ is defined between the pair relative momentum and relative position in the pair rest frame, while $a_{\rm C}$ is the Bohr radius of the pair ($a_C=-83.59$~fm for $\kam p$ pairs). The component $f_{\rm C}(k^*)$ is the strong-interaction scattering amplitude, modified by the Coulomb component:

\begin{equation}
\label{eq:scattering_amplitude}
    f^{-1}_C(k^*) = \frac{1}{f_0} + \frac{1}{2}d_0{\ks}^2-\frac{2}{a_C}h(\ks a_C) -\text{i}\ks a_C,
\end{equation}
where $h(\varepsilon)=\varepsilon^2\sum\limits_{n=1}^{ \infty}[n(n^2+\varepsilon^2)]^{-1}
-\mathrm{{\upgamma}} -\ln|\varepsilon|$
($\mathrm{{\upgamma}} = 0.5772$ is the Euler constant).

Moreover, the description becomes more complicated when coupled channels are present. For details, see Refs.~\cite{Lednicky:2005tb,Haidenbauer:2018jvl}.

The correlation function is obtained by numerical integration of the source function $S(r^*)$, parameterised by a three-dimensional Gaussian in the PRF, with the Bethe--Salpeter amplitude given in Eq.~(\ref{eq:fullpsi2}) and with the Coulomb-modified scattering amplitude defined in Eq.~(\ref{eq:scattering_amplitude})~\cite{Kisiel:2009eh,Kisiel:2018wie}.

% put your appendices here (if any)
%
\end{document}